# Molecular Beam Epitaxy Growth of [CrGe/MnGe/FeGe] Superlattices: Toward Artificial B20 Skyrmion Materials with Tunable Interactions

Adam S. Ahmed,[1] Bryan D. Esser,[2] James Rowland,[1] David W. McComb,[2] and Roland K. Kawakami[1]

[1]Department of Physics, The Ohio State University, Columbus, OH 43210

[2]Center for Electron Microscopy and Analysis, The Ohio State University, Columbus, OH 43212

**Abstract:** Skyrmions are localized magnetic spin textures whose stability has been shown theoretically to depend on material parameters including bulk Dresselhaus spin orbit coupling (SOC), interfacial Rashba SOC, and magnetic anisotropy. Here, we establish the growth of a new class of artificial skyrmion materials, namely B20 superlattices, where these parameters could be systematically tuned. Specifically, we report the successful growth of B20 superlattices comprised of single crystal thin films of FeGe, MnGe, and CrGe on Si(111) substrates. Thin films and superlattices are grown by molecular beam epitaxy and are characterized through a combination of reflection high energy electron diffraction, x-ray diffraction, and cross-sectional scanning transmission electron microscopy (STEM). X-ray energy dispersive spectroscopy (XEDS) distinguishes layers by elemental mapping and indicates good interface quality with relatively low levels of intermixing in the [CrGe/MnGe/FeGe] superlattice. This demonstration of epitaxial, single-crystalline B20 superlattices is a significant advance toward tunable skyrmion systems for fundamental scientific studies and applications in magnetic storage and logic.

e-mail: kawakami.15@osu.edu

Topological spin textures in chiral magnetic materials have become an exciting research direction for condensed matter physics and magnetoelectronic applications[1-9]. In particular, skyrmions are localized spin textures ranging in size from 1 nm to 3 μm, with vortex-like structure as shown in Figure 1. The rotation of magnetization as one circles the core region is characterized by a winding number (= 1 for skyrmions) and non-zero real space Berry curvature. This leads to new phenomena such as the topological Hall effect[7, 10-16] and skyrmion Hall effect[17, 18], as well as the so-called "topological protection" that permits these systems to be robust to perturbations and defects compared to conventional ferromagnetic domain walls[19]. Furthermore, the much lower critical current density necessary for depinning skyrmions (compared to domain walls) makes them promising for potential applications in ultra-high density magnetic memory including racetrack memory[9, 20-25].

Typically, skyrmions are stabilized by the competition between two interactions. The first is the exchange interaction (J), which prefers neighboring spins to be parallel in a ferromagnet. The second competing interaction is the Dzyaloshinskii-Moriya interaction (DMI), which prefers neighboring spins to be 90° to one another. The DMI can only appear in systems that break bulk inversion symmetry or mirror symmetry. The first experimental evidence for the existence of skyrmions in condensed matter systems was shown in bulk MnSi crystals[26], where the bulk inversion symmetry breaking of its B20 crystal structure generates a Dresselhaus spin orbit coupling and produces a non-zero DMI ($D_D$). Subsequent efforts continued on a variety of bulk B20 skyrmion crystals including FeGe[23, 27-32]. With the development of B20 thin film systems, skyrmion stability was dramatically increased both in magnetic field and temperature range compared to their bulk counterparts[31]. For example, FeGe bulk crystals stabilize the skyrmion phase in a small temperature window of a few Kelvin below the ordering temperature (280 K). In contrast, sputter-deposited FeGe thin films show a dramatic extension of temperature stability from 280 K down to 10 K measured from topological Hall resistivity[27]. Moreover, it was shown that these same sputtered films had a finite zero-field topological Hall signal—an indication of zero field skyrmions[27, 33]. These were important indications that film thickness and magnetic anisotropy could impact the skyrmion phase diagram in favorable ways for enhanced skyrmion stability.

Motivated by these experiments, Randeria and co-workers[34, 35] modeled the skyrmion phase diagram for 2D systems as a function of magnetic anisotropy ($K < 0$ for out-of-plane hard axis, or "easy plane"; $K > 0$ for out-of-plane easy axis, or "easy axis"). 2D systems, such as thin films and interfaces, break mirror symmetry which generates a Rashba spin orbit coupling and produces a DMI ($D_R$) which has a different functional form than the Dresselhaus DMI, $D_D$[35]. Notably, the theoretical calculations have shown a vastly increased region of skyrmion stability when $D_R > D_D$ and the magnetic anisotropy is easy plane. Currently, experimental work has been limited to either the Dresselhaus or the Rashba limit where skyrmions are either vortex-like Bloch skyrmions[14, 27, 29, 31, 36-40] (e.g. FeGe, MnSi, $Fe_{0.8}Co_{0.2}Si$) or hedgehog-like Néel skyrmions[3, 41] (e.g. Pt/Co/MgO, Ta/$Co_{20}Fe_{60}B_{20}$/$TaO_x$), respectively (Figure 1). Developing systems that can continuously tune skyrmions from Bloch to Néel can help realize optimal conditions for skyrmion stability and device operation (e.g. pinning potentials, skyrmion velocity, topological Hall effect). Therefore, in order to explore the predictions of Randeria and co-workers, it is crucial to develop magnetic materials where the parameters $D_D$, $D_R$, and K can be tuned independently.

To this end, one may consider developing heterostructures of B20 thin films. For a single interface between two materials, a non-zero Rashba DMI can develop due to the broken mirror symmetry. In a 2-component [A/B]$_N$ superlattice (with N repetitions), mirror symmetry remains intact and the Rashba DMI from the A/B interface cancels with the B/A interface. However, the magnetic anisotropy K from the interfaces is additive and does not cancel. In contrast, a 3-component superlattice [A/B/C]$_N$ can yield a non-zero overall $D_R$ because the A/B, B/C, and C/A interfacial contributions to the DMI need not cancel. Because $D_R$ originates from the interfaces while $D_D$ is from the bulk, the relative strengths can be tuned by layer thickness with larger $D_R/D_D$ for thinner layers. Based on these considerations, it will be possible to independently tune the overall $D_D$, $D_R$, and K of the system through precise growth of B20 superlattices and heterostructures. However, to our knowledge, such artificial B20 structures have yet to be synthesized.

In this manuscript, we demonstrate the growth of B20 superlattices by molecular beam epitaxy (MBE), which establishes a route toward tunable skyrmion materials. We begin by growing FeGe(111) films on Si(111) substrates to establish a single-crystalline B20 template. We follow this growth with MnGe/FeGe and CrGe/FeGe bilayers to check for good epitaxial registry and favorable growth conditions between different B20 materials. We characterize all thin films to ensure single crystalline quality with reflection high energy electron diffraction (RHEED) and x-ray diffraction (XRD). Then, two-component [A/B]$_N$ superlattices of [MnGe/FeGe]$_8$ and [CrGe/FeGe]$_{10}$ are synthesized and characterized with XRD and RHEED. Finally, a three-component superlattice of [CrGe/MnGe/FeGe]$_8$ is grown and further characterized by cross-sectional scanning transmission electron microscopy (STEM) and x-ray energy dispersive spectroscopy (XEDS) to reveal chemical specificity and atomic layer resolution. This set of novel heterostructures demonstrates a fundamental materials advance for tunable skyrmions in artificial B20 systems.

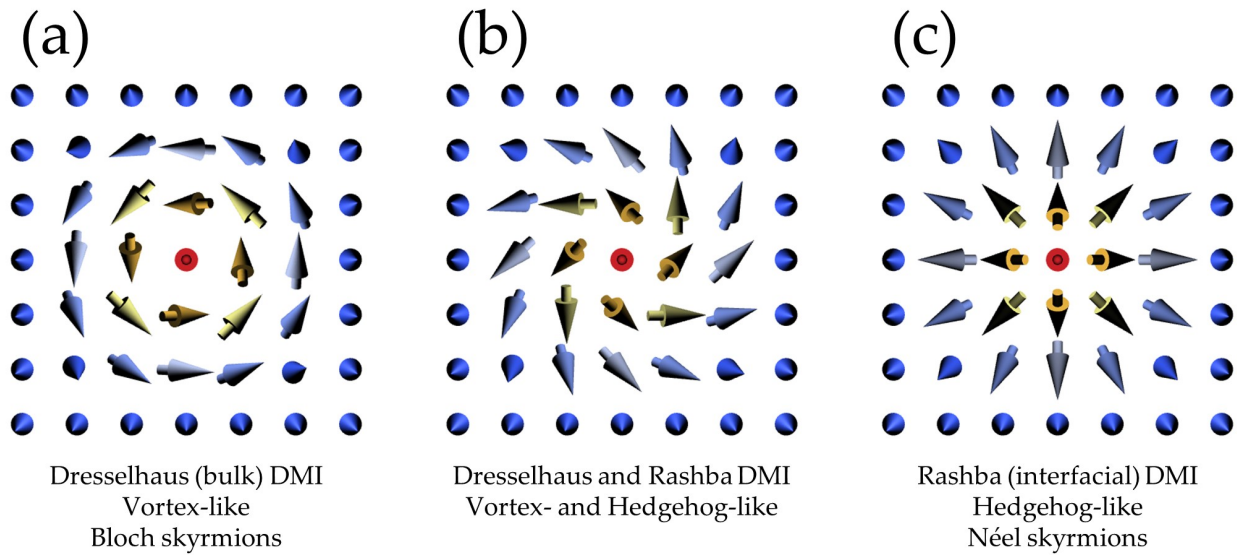

**Figure 1.** Diagram illustrating skyrmion spin textures tuning the relative ratio of $D_D/D_R$. (a) Dresselhaus limit ($D_R/D_D \rightarrow 0$): vortex-like (Bloch) skyrmions characteristically appear in B20 materials where only bulk inversion symmetry is broken. (b) Intermediate regime ($D_R/D_D \sim 1$): spin textures share qualities of both Bloch and Néel skyrmions in systems with both Dresselhaus and Rashba DMI; these skyrmions have yet to be observed. (C) Rashba limit ($D_R/D_D \rightarrow \infty$): hedgehog-like (Néel) skyrmions are present in magnetic multilayers where only surface inversion symmetry is broken.

**Experimental Methods**

The B20 thin films were grown by MBE in an ultrahigh vacuum (UHV) chamber with a base pressure of $2 \times 10^{-10}$ torr. Si(111) substrates (MTI, resistivity 10,000 $\Omega$-cm, single-side polished) were prepared by first sonicating the wafers in acetone for 5 minutes and isopropyl acetone for 5 minutes to remove any loose dirt/particles prior to treatment. The cleaned wafers were then dipped in buffered HF (Alfa Aesar) for 2 minutes to remove the native oxide and terminate the dangling Si bonds with H. The substrates were quickly loaded into the growth chamber and pre-annealed at 800 °C for 20 minutes to desorb the hydrogen and simultaneously obtain a 7x7 reconstruction. A quartz deposition monitor was utilized to determine the approximate thicknesses of the films. For the growth of FeGe films and buffer layers, the substrate temperature was held at 300 °C. Elemental Fe and Ge were flux matched in a 1:1 ratio and co-deposited on the Si(111) substrate. For growth of MnGe and CrGe (flux matched 1:1) and all superlattices, the substrate temperature was maintained at 250 °C. Samples were characterized with *in situ* RHEED to obtain qualitative information about in-plane crystallinity. Out-of-plane lattice constants were quantified with x-ray diffraction (XRD) for all samples. For the superlattice, we performed cross-sectional STEM and XEDS for compositional mapping of individual layers.

**FeGe thin films**

FeGe is a magnetic material with B20 crystal structure and is known to host the skyrmion phase[23, 27-29, 31, 32]. The helical pitch length is approximately 70 nm, and the skyrmions form an incommensurate hexagonal lattice. It has an ordering temperature ($T_C$) of 280 K, which is the highest known value for a B20 skyrmion materials. Its bulk cubic lattice constant[30] is 4.679 Å. For FeGe films with (111) orientation, this corresponds to an out-of-plane lattice constant of 8.104 Å = $\sqrt{3}$ (4.679 Å) and a quadruple-layer (QL) period of 2.701 Å, as shown in Figure 2a (a QL consists four atomic layers: dense Fe, dense Ge, sparse Fe, sparse Ge). Of the three B20 materials presented in this manuscript, FeGe is the most well-known and studied.

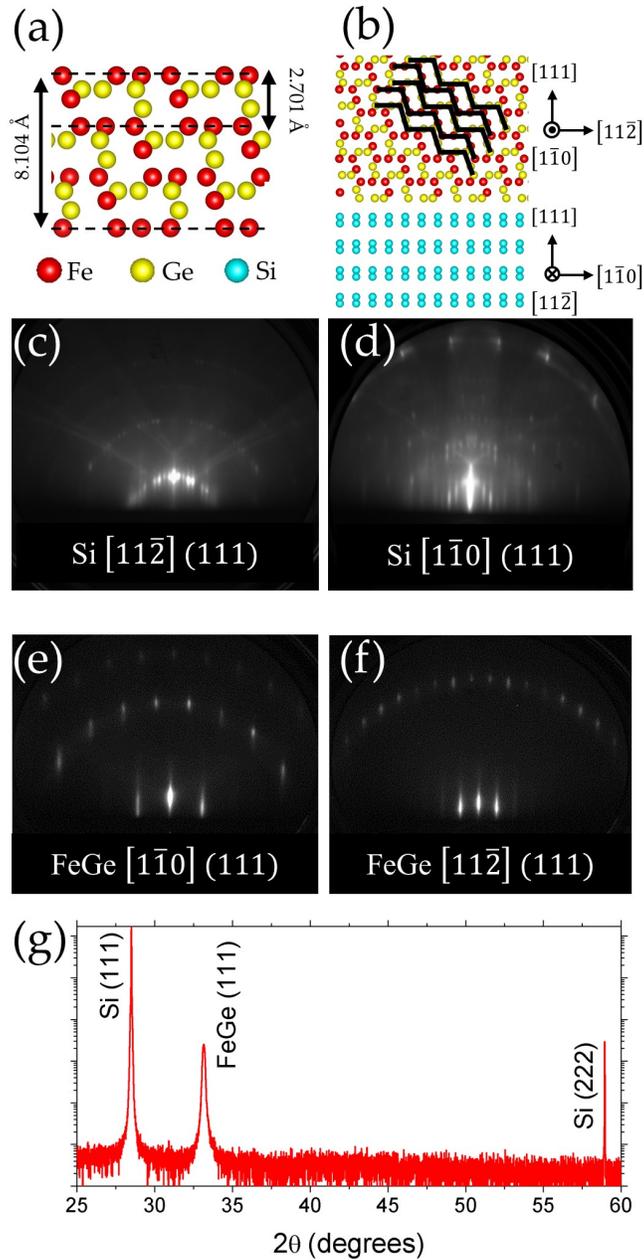

**Figure 2.** (a) Cross-sectional diagram of FeGe showing the (111) lattice constant of 8.104 Å and the quadruple layer (QL) period of 2.701 Å. (b) Diagram of the epitaxial alignment of FeGe(111) on Si(111) with 30° lattice rotation. (c, d) RHEED images of a 7x7 reconstructed Si(111) surface shown for in-plane directions of [11$\bar{2}$] and [1$\bar{1}$0], respectively. (e, f) RHEED patterns measured along the [1$\bar{1}$0] and [11$\bar{2}$] in-plane directions of the ~40 nm FeGe film, respectively. (g) XRD scan of the FeGe film. Single-phase, single crystal FeGe is shown amongst the Si substrate peaks.

We first established the growth of FeGe thin films on Si(111) substrates. RHEED patterns of the Si(111) surface taken along the [11$\bar{2}$] (Figure 2c) and [1$\bar{1}$0] (Figure 2d) in-plane directions show a clear 7x7 reconstruction pattern after the 800 °C anneal. This indicates a clean and well-ordered surface for subsequent growth of B20 materials. For growth of FeGe in the B20 phase, the substrate temperature was reduced to 300 °C, and elemental sources of Fe and Ge were co-deposited with a flux ratio of 1:1. The pressure during growth was 1x10$^{-9}$ torr. Figures 2e and 2f show RHEED patterns for a ~40 nm FeGe film. Notably, the FeGe lattice is rotated relative to the Si(111) lattice in order to minimize the lattice mismatch, thereby having the [1$\bar{1}$0] axes of FeGe align with the [11$\bar{2}$] axes of Si(111), and the [11$\bar{2}$] axes of FeGe align with the [1$\bar{1}$0] axes of Si(111). This epitaxial alignment is illustrated in the crystal structure diagrams in Figure 2b. The sharp streaks in the RHEED pattern are indicative of flat terraces, and the distinct patterns along the [1$\bar{1}$0] (Figure 2e) and [11$\bar{2}$] (Figure 2f) directions of FeGe indicate in-plane single crystal order.

To further investigate the crystallinity and orientation of the FeGe films, we performed XRD in the θ-2θ geometry. Figure 2g shows an XRD scan for the ~40 nm FeGe/Si(111) sample. The Si(111) and Si(222) peaks are present from the substrate. Additionally, an FeGe(111) peak appears at 33.10° and no other phases of Fe:Ge are present, showing that single phase FeGe was synthesized. From this, we calculated the QL period to be 2.703 Å, which is larger than the bulk value by ~0.065%. This difference can be attributed to in-plane compressive strain on the FeGe film due to the 0.05% lattice mismatch between FeGe(111) and Si(111).

Having established the single crystal growth of FeGe films on Si(111), we utilized this as a template for subsequent growth of B20 overlayers and superlattices.

**MnGe thin films and superlattices**

MnGe is another skyrmion B20 crystal material. The helical pitch length is ~3 nm and forms a square lattice of skyrmions[42]. It has an ordering temperature of 170 K, and shows the largest topological Hall effect of the B20 compounds[11]. The bulk lattice constant[43] is 4.795 Å, which corresponds to a QL period of 2.768 Å.

We first established the growth of MnGe thin films on FeGe/Si(111) to determine suitable growth conditions for synthesizing MnGe/FeGe superlattices. Starting with a ~5 nm FeGe base layer grown on Si(111), a ~40 nm MnGe overlayer was co-deposited from elemental sources of Mn and Ge. The flux ratio was held at 1:1 and the substrate temperature was maintained at 250 °C during growth. Figures 3a and 3b show the RHEED images for the MnGe film along the $[1\bar{1}0]$ and $[11\bar{2}]$ directions, respectively. The RHEED images are streaky, indicative of flat terrace growth on the FeGe base layer. To quantify the out-of-plane lattice constant, Figure 3c shows XRD for the MnGe/FeGe/Si(111) bilayer. First, there are no other phases of $Mn_xGe_y$ present other than MnGe. Second, a MnGe(111) peak appears at a 2θ of 32.41°. This corresponds to a QL period of 2.759 Å, which is smaller than the bulk value by 0.33%. Comparing this value to the 2.5% lattice mismatch between MnGe and FeGe suggests that the ~40 nm MnGe film has relaxed to its bulk-like lattice structure. There is an additional shoulder peak which can be attributed to the ~5 nm FeGe base layer. The RHEED combined with XRD data shows that MnGe grows single crystalline on the FeGe(111) template.

Having found suitable growth conditions for MnGe/FeGe/Si(111) bilayers, we subsequently demonstrated the ability to synthesize a two-component superlattice comprised of MnGe (~2 nm) and FeGe (~2 nm). The first layer grown was FeGe/Si(111) where the substrate temperature was held at 300 °C. After we obtained a good RHEED image of FeGe, the substrate was cooled to 250 °C for the remainder of the superlattice growth. Figures 3e and 3f show the RHEED for the final FeGe and MnGe layers, respectively. Both images are qualitatively very similar, and during growth, there was very little change in the RHEED images for the different layers. Figure 3d shows the XRD scan of the [MnGe/FeGe]$_8$/Si(111) superlattice. Instead of two peaks for MnGe and FeGe, a single peak is present at a 2θ of 32.72°, which corresponds to a QL period of 2.734 Å, which lies between the measured QL periods of thick films of FeGe (2.703 Å) and MnGe (2.759 Å). Looking more closely near this peak reveals the presence of a weak satellite peak as indicated by the arrow in the inset of Figure 3d.

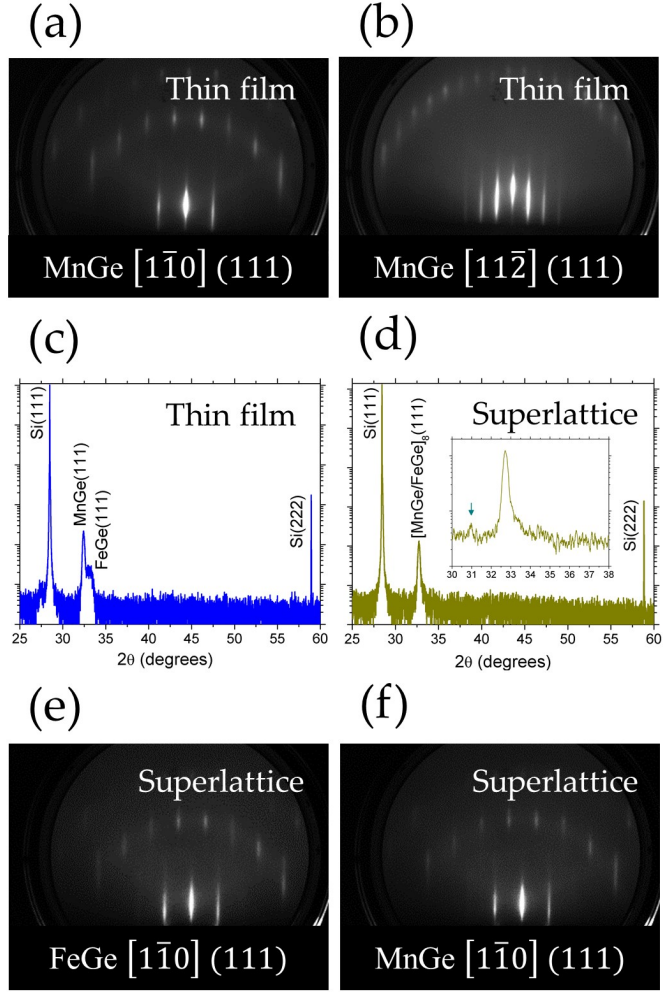

**Figure 3.** *Thin film:* RHEED for MnGe thin film grown on FeGe buffer layers is shown along the (a) $[1\bar{1}0]$ and (b) $[11\bar{2}]$ directions. (c) XRD for a ~40 nm thick MnGe film. The MnGe(111) peak is shown alongside the ~5 nm FeGe buffer layer. *Superlattice:* (d) XRD for the superlattice structure shows a single (111) peak. Inset: a weak satellite peak (arrow) due to superlattice structure is observed. (e, f) RHEED patterns are shown for the final topmost layers in a [MnGe/FeGe]$_8$ superlattice for FeGe$[1\bar{1}0]$(111) and CrGe $[1\bar{1}0]$(111), respectively.

## **CrGe thin films and superlattices**

In addition to magnetic B20 materials, we are also interested in incorporating nonmagnetic materials. CrGe is a B20 paramagnetic metal that does not have a magnetic ordering temperature and, consequently, is not known to host a skyrmion phase[44-47]. Its bulk lattice parameter is 4.790 Å (QL period of 2.766 Å) and is very closely latticed matched to MnGe. To the best of our knowledge, the growth of CrGe thin films has yet to be demonstrated.

Prior to growth of CrGe, a ~5 nm FeGe base layer was grown on a 7x7 reconstructed Si(111) substrate at 300 °C. Elemental sources of Cr and Ge were co-deposited on the FeGe base layer at 250 °C. Figures 4a and 4b show the CrGe RHEED images for a ~30 nm thick film. The RHEED images are qualitatively similar to MnGe and FeGe, but the CrGe streaks have bright spots and arrowhead-like features which indicate the onset of faceting and 3D growth[48]. In spite of these features, XRD was performed to obtain the out-of-plane CrGe lattice constant (Figure 4c). In addition to the FeGe(111) peak from the buffer layer, a CrGe(111) peak appears at a 2θ of 32.38°, corresponding to a QL period of 2.761 Å. This is smaller than the bulk value by only 0.15%, which suggests that the ~30 nm CrGe film has relaxed to its bulk-like lattice structure.

Having found suitable growth conditions for CrGe thin films, we applied these similar growth parameters to the fabrication of a two-component superlattice. A [CrGe/FeGe]$_{10}$ superlattice was grown on Si(111) at 250 °C with each layer having a thickness of approximately 2 nm. The RHEED patterns during superlattice growth are shown in Figures 4e and 4f. In contrast to the growth of the ~30 nm CrGe film, the RHEED is streaky and absent of any arrowhead features. Lastly, Figure 4d shows an XRD scan of the superlattice structure, and a [CrGe/FeGe]$_{10}$(111) peak is located at 2θ of 32.68°, corresponding to a QL period of 2.737 Å. A closer look near this peak clearly shows the presence of additional satellite peaks due to the superlattice structure (Figure 4d inset). The stronger satellite peaks for CrGe compared to MnGe suggest sharper interfaces. The spacing of the satellite peaks corresponds to a superlattice period of ~5 nm, which is consistent with the designed period of ~4 nm. With this, we have established that B20 CrGe layers are suitable for artificial B20 heterostructures.

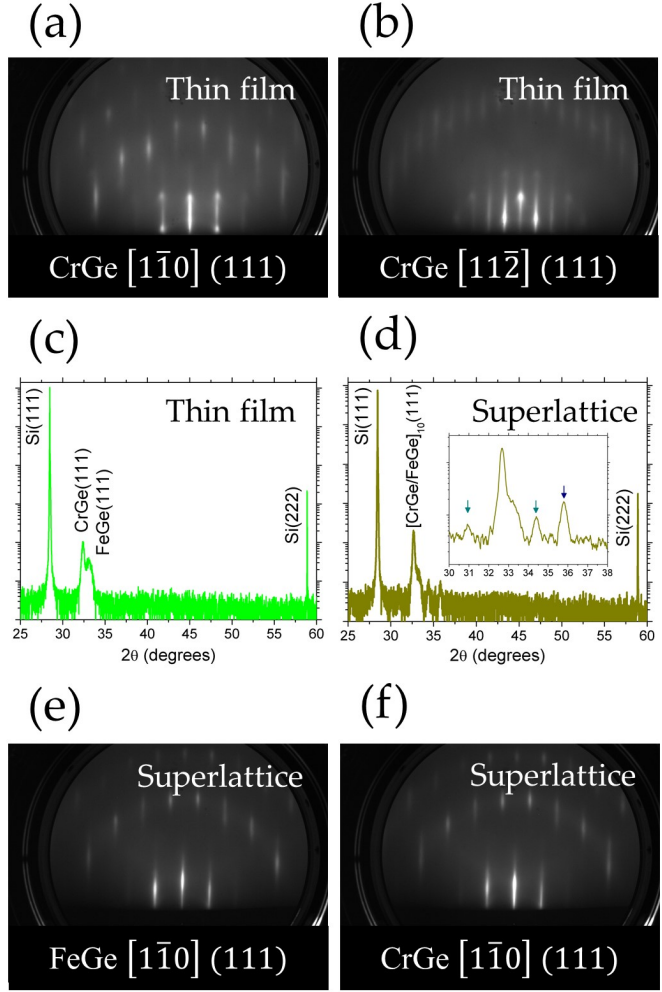

**Figure 4.** *Thin film:* RHEED for CrGe thin film grown on FeGe buffer layers is shown along the (a) [1$\bar{1}$0] and (b) [11$\bar{2}$] directions. Arrowhead features indicate the occurrence of faceting. (c) XRD for a ~30 nm thick CrGe film. The CrGe(111) peak is shown alongside the ~5 nm FeGe buffer layer. *Superlattice:* (d) XRD for the superlattice structure shows a single (111) peak. Inset: satellite peaks (arrows) are observed due to the superlattice structure. (e, f) RHEED patterns are shown for the final topmost layers in a [CrGe/FeGe]$_{10}$ superlattice for FeGe[1$\bar{1}$0](111) and CrGe [1$\bar{1}$0](111), respectively.

## [CrGe/MnGe/FeGe] trilayer superlattice

With suitable growth conditions obtained for CrGe, MnGe, and FeGe, we proceeded to the synthesis of a three-component B20 superlattice. A [CrGe/MnGe/FeGe]$_8$ superlattice structure was grown at 250 °C on a 7x7 reconstructed Si(111) surface. Every layer had a target thickness of approximately 2 nm. Figures 5a, 5b, and 5c show RHEED patterns in the [1$\bar{1}$0] direction for the topmost FeGe, MnGe, and CrGe layers in the superlattice, respectively. The RHEED is streaky indicating smooth 2D surfaces, and all three layers have similar patterns. Figure 5d shows the XRD scan, which exhibits a [CrGe/MnGe/FeGe]$_8$(111) peak at 2θ of 32.65° corresponding to a QL period of 2.740 Å. The inset of Figure 5d shows the presence of satellite peaks corresponding to a superlattice period of ~7 nm, which is consistent with the designed period of ~6 nm.

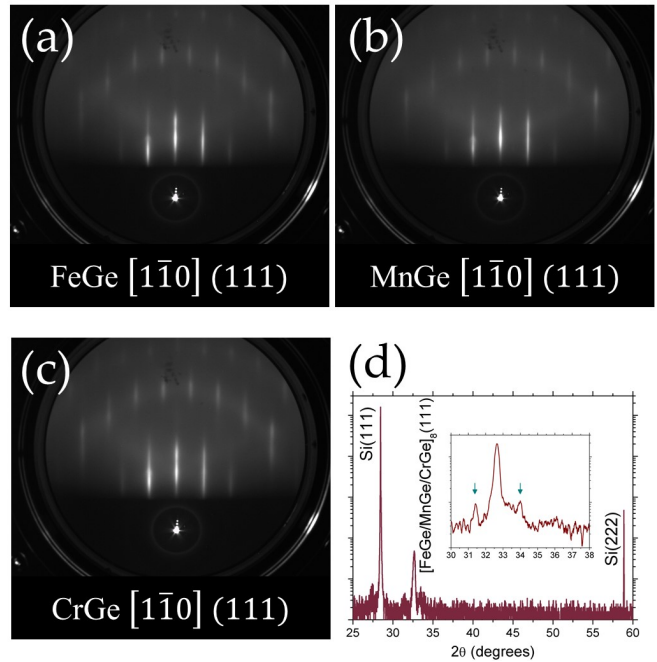

**Figure 5.** RHEED and XRD characterization of a three-component superlattice [CrGe/MnGe/FeGe]$_8$. (a-c) RHEED patterns are shown for the topmost layers FeGe, MnGe, and CrGe layers, respectively. The patterns are measured along the [1$\bar{1}$0] in-plane direction of the films. Qualitatively, the RHEED images are very similar between each layer. (d) XRD scan of the trilayer superlattice shows a single (111) peak with satellite peaks from the superlattice structure (arrows in the inset).

To gain further insight into the structural and compositional quality of the three-component superlattice structure, cross-sectional STEM and XEDS were performed on an aberration corrected FEI Titan³ G2 60-300 S/TEM equipped with a four quadrant Super-X XEDS detection system. A STEM image and XEDS compositional maps acquired from a cross-section with the viewing direction parallel to the FeGe [1$\bar{1}$0] axis are shown in Figure 6. The contrast in the high-angle annular dark field (HAADF) STEM image (Figure 6a) arises from differences in atomic number (Z), where heavier elements have higher intensity. The compositional maps were extracted from the hyperspectral XEDS dataset. The uniform intensity in the Ge map (Figure 6b) demonstrates the high degree of Ge homogeneity, which is consistent with the stoichiometry of the B20 materials grown. Figures 6d, 6e, and 6f are the elemental maps for Fe, Mn, and Cr, respectively, which show the quality of the MBE growth process with consistent, high quality layers with relatively low interdiffusion. The composite elemental XEDS map of Fe, Mn, and Cr (Figure 6c) shows the relative positions of the layers and provides further confirmation of the quality of the multilayer film. The high resolution HAADF-STEM image in Figure 7 demonstrates the high quality single crystalline layers of the superlattice, as evidenced by the coherent interfaces and block-like, zig-zag structure characteristic of B20 crystals along $[1\bar{1}0]$ (Figure 2b).

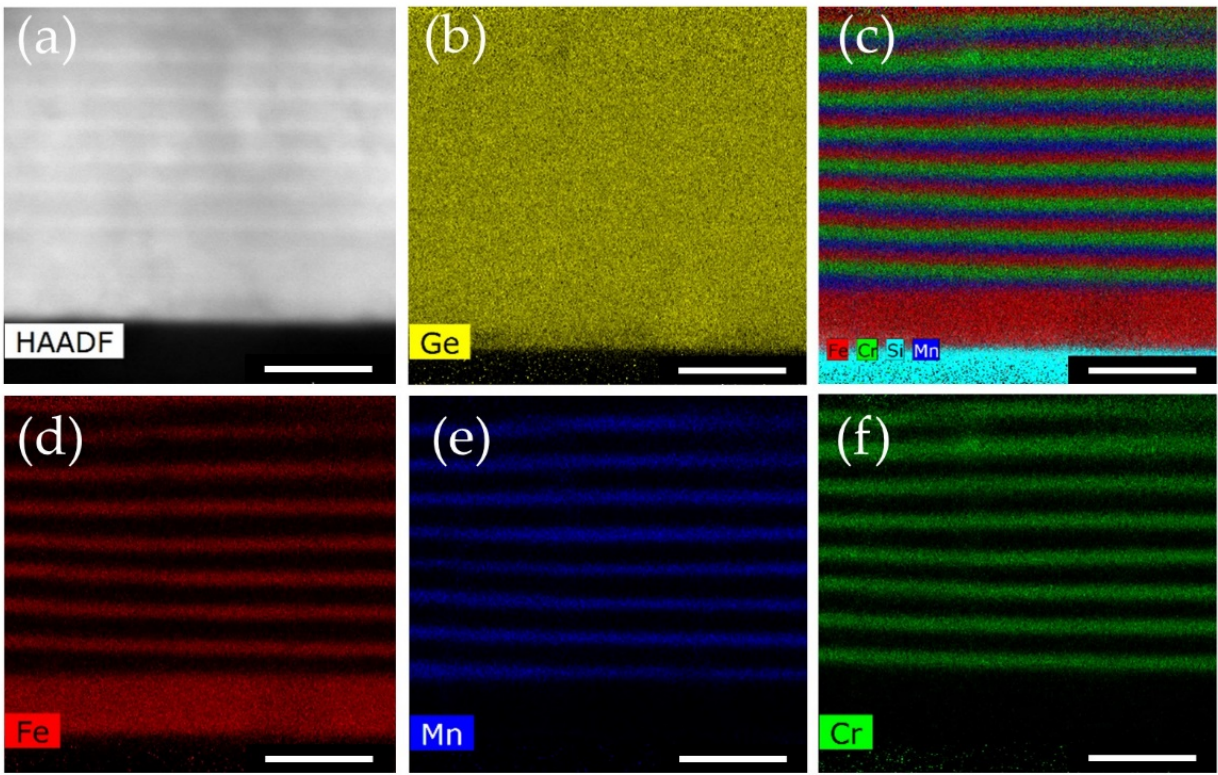

**Figure 6.** Cross-sectional STEM XEDS for the [CrGe/MnGe/FeGe]$_8$ superlattice. (a) HAADF image shows initial contrast for superlattice structure. (b) An elemental map for Ge. (c) A composite elemental map showing Fe (red), Cr (green), Si (light blue), and Mn (blue). (d-f) Separate elemental maps for Fe, Mn, and Cr, respectively. All scale bars are 20 nm.

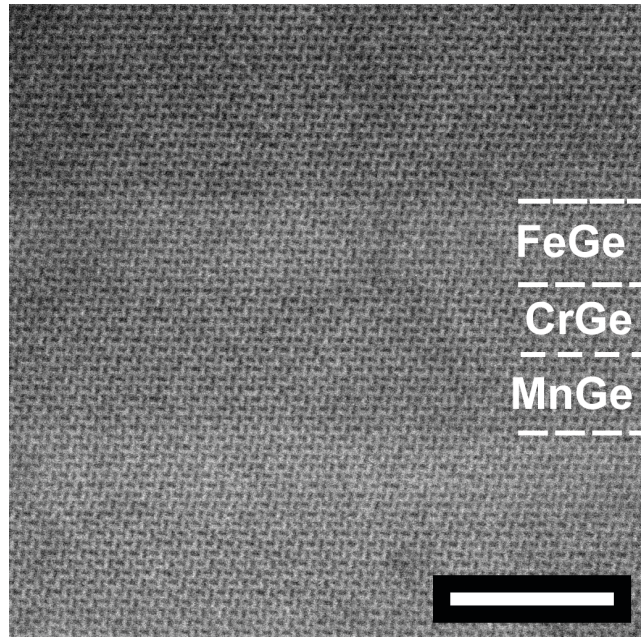

**Figure 7.** HAADF-STEM image of the [CrGe/MnGe/FeGe]$_8$ superlattice showing completely coherent interfaces between each layer and clear B20 structure, as indicated by the zig-zag pattern described in Figure 2b. Z-contrast and the XEDS data distinguish between each layer. The scale bar is 5 nm.

## Conclusion

We have synthesized the first B20 superlattices by MBE, including two-component superlattices ([MnGe/FeGe] and [CrGe/FeGe]) and a three-component superlattice ([CrGe/MnGe/FeGe]). Characterization by RHEED, XRD, and cross-sectional STEM indicate high crystalline quality and relatively low elemental interdiffusion at the interfaces. These studies establish the growth of a new class of artificial materials that offer a path toward skyrmion systems with tunable $D_D$, $D_R$, and K for fundamental scientific studies and applications in magnetic memory and logic.

## Acknowledgements

ASA and RKK acknowledge support from the Ohio State Materials Seed Grant (MTB-G00010). BDE and DWM acknowledge support from the Center for Emergent Materials at the Ohio State University, a National Science Foundation Materials Research Science and Engineering Center (Grant No. DMR-1420451), as well as the Ohio State Materials Seed Grant (MTB-G00012) and partial support from the Center for Electron Microscopy and Analysis. JR acknowledges support from the NSF graduate fellowship.